\documentclass[
]{ceurart}

\usepackage{listings}
\usepackage{tcolorbox}
\usepackage{graphicx}
\usepackage{subcaption}
\begin{document}

\copyrightyear{2026}

\copyrightclause{Copyright for this paper by its authors. Use permitted under Creative Commons License Attribution 4.0 International (CC BY 4.0).}
\conference{The 3rd Solid Symposium Poster Session, co-located with the 4th Solid Symposium, April 30 - May 01 2026, London, UK}

\title{LODox: A Web-Based Platform for Cultural Heritage Linked Data Discovery}


\author[1]{Puyu Wang}[%
orcid=0009-0003-8089-4146,
email=puyu.wang@eng.ox.ac.uk,
url=https://puyuwang.org/,
]
\cormark[1]
\address[1]{Oxford e-Research Centre, Department of Engineering Science, University of Oxford, UK}

\author[2]{Rui Zhao}[%
orcid=0000-0003-2993-2023 ,
email=zhaorui@tju.edu.cn ,
url=https://me.ryey.icu,
]
\address[2]{School of Cybersecurity, Tianjin University, China}
\author[3]{Hang Dong}[%
orcid=0000-0001-6828-6891,
email=h.dong2@exeter.ac.uk,
url=https://experts.exeter.ac.uk/41644-hang-dong,
]
\address[3]{Department of Computer Science, University of Exeter, UK}

\cortext[1]{Corresponding author.}

\begin{abstract}
Recent advances in Large Language Models have transformed how data can be linked, discovered, and enhanced. LODox, Linked Open Data at Oxford, is an evolving web-based platform that embraces this transformation for cultural heritage Linked Open Data. LODox  features a novel interface with intuitive card-based data presentation and visualization in a semantic portal. It represents an evolving approach to explore heterogeneous cultural heritage data across multiple sources through a unified interface. In this paper, we will briefly describe the LODox architecture, its implementation, and future development plan, particularly its alignment and integration with Solid. 

\end{abstract}

\begin{keywords}
  Data Discovery \sep
  Linked Data \sep
  Cultural Heritage \sep
  Semantic Portal 
\end{keywords}

\maketitle

\section{Introduction}

The cultural heritage domain faces long-standing challenges in accessing and integrating Linked Data, particularly the technical barriers from the complexity of query languages and ontologies, underscoring the need for robust, user-friendly systems designed for the domain. 
To address these barriers, a range of semantic portals and Linked Data platforms have been developed, from early SPARQL-based systems for scholarly data exploration \cite{hutchison_linked-data-driven_2013} to more comprehensive frameworks such as ResearchSpace, Omeka S\footnote{https://omeka.org/}, and Sampo-UI \cite{haase_metaphactory_2019, vrandecic_reshaping_2018, ikkala_sampo-ui_2021, hyvonen_digital_2023, hyvonen_how_2025, miranda_cidoc-crm_2025}. These systems support activities including semantic search, knowledge graph construction, Linked Data publication, and rapid development of domain-specific portals, demonstrating substantial progress toward making cultural heritage data more accessible and interoperable.

However, existing solutions continue to present challenges for both non-technical researchers and technical developers. In particular, effective use of standards such as CIDOC-CRM \footnote{https://cidoc-crm.org/} and Linked Data technologies often requires considerable expertise in ontology modelling and semantic data management \cite{miranda_cidoc-crm_2025}.
Large Language Models (LLMs) have demonstrated human-like capabilities in various tasks. For example, query generation from natural language is an important step in Retrieval-Augmented Generation (RAG) with knowledge graph pipeline \cite{diallo_comprehensive_2024, fondi_integrating_2024, hoseini_towards_2024}. LLMs offer a promising route to lower the technical barriers\cite{minaee_large_2025}; integrating LLM into Linked Data workflows and platforms is, therefore, an increasingly important direction. 
However, there might be considerable re-engineering needed to integrate LLMs into existing cultural heritage platforms. This motivates a lightweight, modular platform, designed to make integrating LLMs components easier. To this end, we present LODox, a prototype web-based platform built with Django with minimal stack for exploring and visualizing Linked Open Data. We consider such minimal stack should include the following essential components:

\begin{itemize}
    \item \textbf{Database}: database system for persistent storage of graph data, for example, blazegraph \footnote{https://blazegraph.com/}. 
    \item \textbf{Resolution system}: resolution system that is able to resolve identifiers (e.g. URIs) to corresponding resources.
    \item \textbf{Query Interface}: interface for data query or manipulation, for example, a SPARQL endpoint.
    \item \textbf{Graphical User Interface}: LOD visualisation tool that serves an entry point for querying, discovering, and interaction with interconnected datasets.
\end{itemize}
LODox is built on a state-of-the-art web stack, and its modular architecture facilitates straightforward integration of LLM pipelines. Furthermore, we envision LODox as a pathway towards decentralised community-enriched data curation and discovery with Solid protocol \cite{sambra_solid_2016}. This paper will present and discuss the architecture of the LODox system, the implementation, and outlines this pathway.

\section{System Architecture of LODox}
\subsection{High-level Overview}

\begin{figure}[ht]
    \centering
    \includegraphics[width=0.8\linewidth]{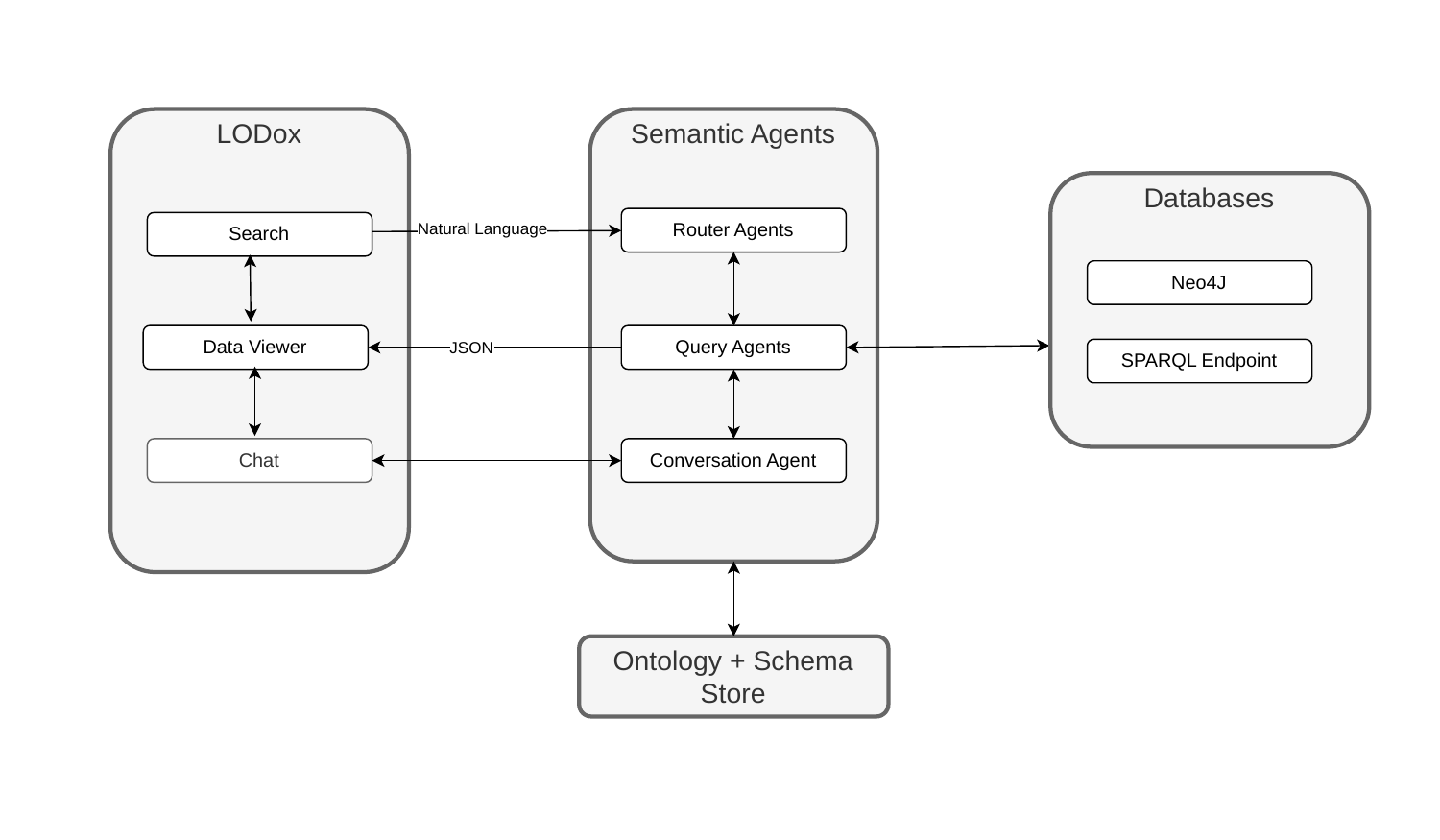}
    \caption{A High-level Overview of System Architecture}
    \label{fig:stackio}
\end{figure}

As shown in Figure~\ref{fig:stackio}, the LODox user interface have various components, including Search and Data Viewer. Users start with LODox search; results are visualized in the LODox Data Viewer. The \emph{Semantic Agents} illustrate how LODox can be integrated with LLM-based agents; they are invocable as Python packages within the backend, and can be bypassed for certain purposes (\textit{e.g.}, for reproducibility testing or debugging). This decoupled design provides several advantages: (1) components can be updated and tested independently without affecting one another, allowing more rapid development cycles, (2) they could be deployed or used independently, (3) the modular design provides the flexibility to build more modules for LODox in the future. 
\subsection{Implementation of LODox}
\begin{figure*}[ht]
    \centering
    \includegraphics[width=0.8\linewidth]{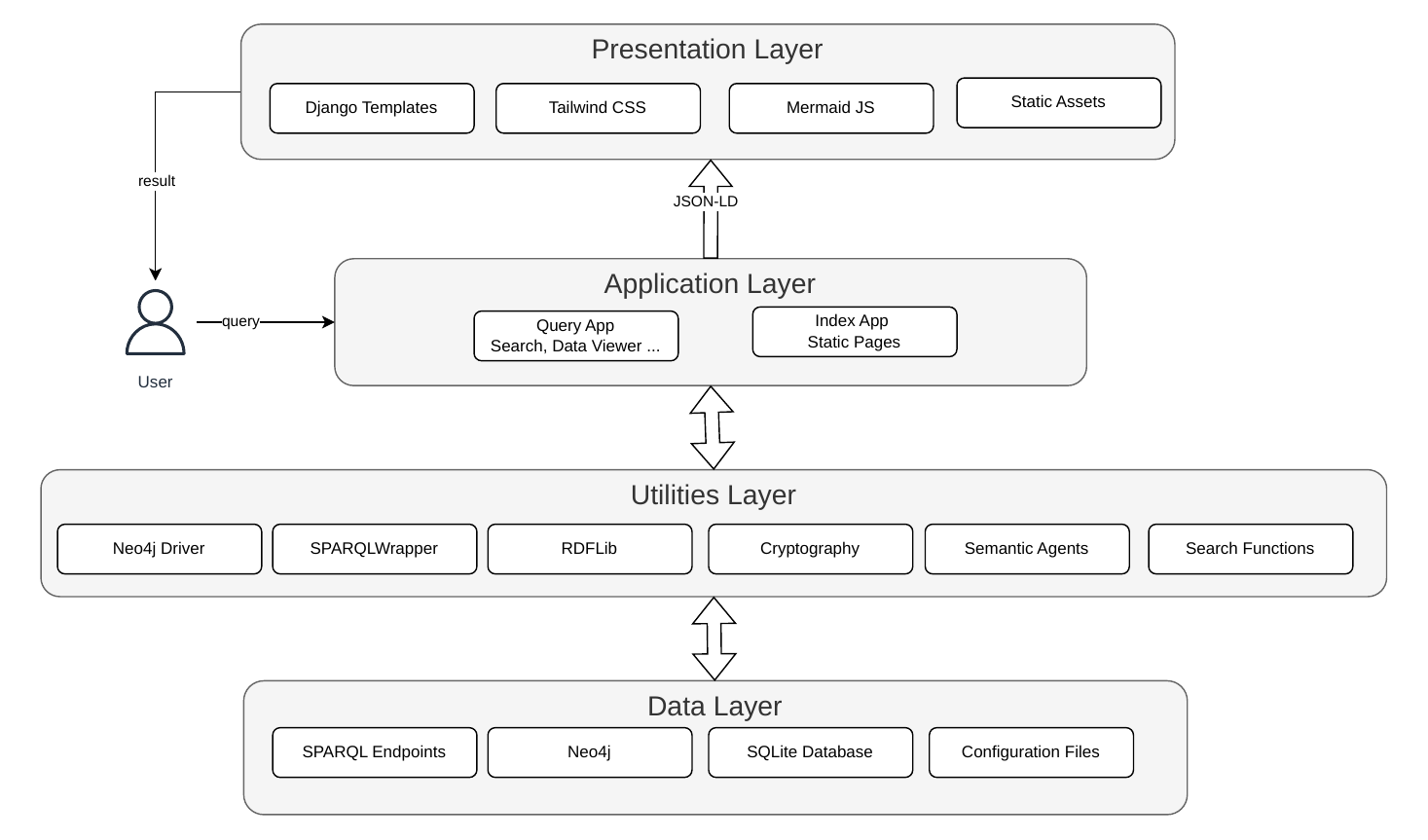}
    \caption{LODox Layers}
    \label{fig:LODoxLayers}
\end{figure*}

Within the LODox, the system can be understood through the layers as shown in  Figure ~\ref{fig:LODoxLayers}. There are four layers:

1. \textbf{Presentation Layer}. The presentation layer is where users interact with the system and explore the data. It is built from reusable templates that are dynamically composed and rendered by Django Framework.  
We choose the Tailwind CSS\footnote{https://tailwindcss.com/} as CSS library considering the its widespread adoption in the frontend community, strong community support, and production maturity. We choose Mermaid JS \footnote{https://mermaid.js.org/} as the diagramming tool over the D3 JS \footnote{https://d3js.org/} considering it provides a declarative diagram language for drawing diagram, which allows easier integration with LLMs. The static assets provides images and other supporting resources required for the system runtime. 


2. \textbf{Application Layer}. The application layer includes two Django Apps in the system: Query App and Index App (the default app). The Index app serves a statics page introducing the system. The Query App orchestrates user interactions, handle application logic, and enforce business rules.
The Query App will route the user's request to  the corresponding functions or utilities. For resource detail requests, the Query App will call Data Viewer page and retrieve the appropriate resource(s) and then send the data, in JSON-LD format, to presentations layer. 

3. \textbf{Utilities Layer}. Utilities layer consists of tools and third-party libraries that can be invoked from the application layer, including the database drivers, search functions, and (where applicable), JSON-LD parsers. Some other tools, omitted from the figure, serve the system itself, for example, loading layout settings. 

4. \textbf{Data Layer}. SQlite, the default database in Django, is used for persisting application data, such as endpoint configurations and associated credentials. It is used for local persistence of application data \textit{(e.g.}, endpoint configurations and associated credentials). Credentials are encrypted and can be accessible only to authenticated users. A user interface is provided for endpoint management. The utilities layer sends queries to the endpoints and returns the results to application layer and then to presentation layer. 
\subsection{Card-Based Data Viewer System}
\begin{figure}
    \centering
    \includegraphics[width=1\linewidth]{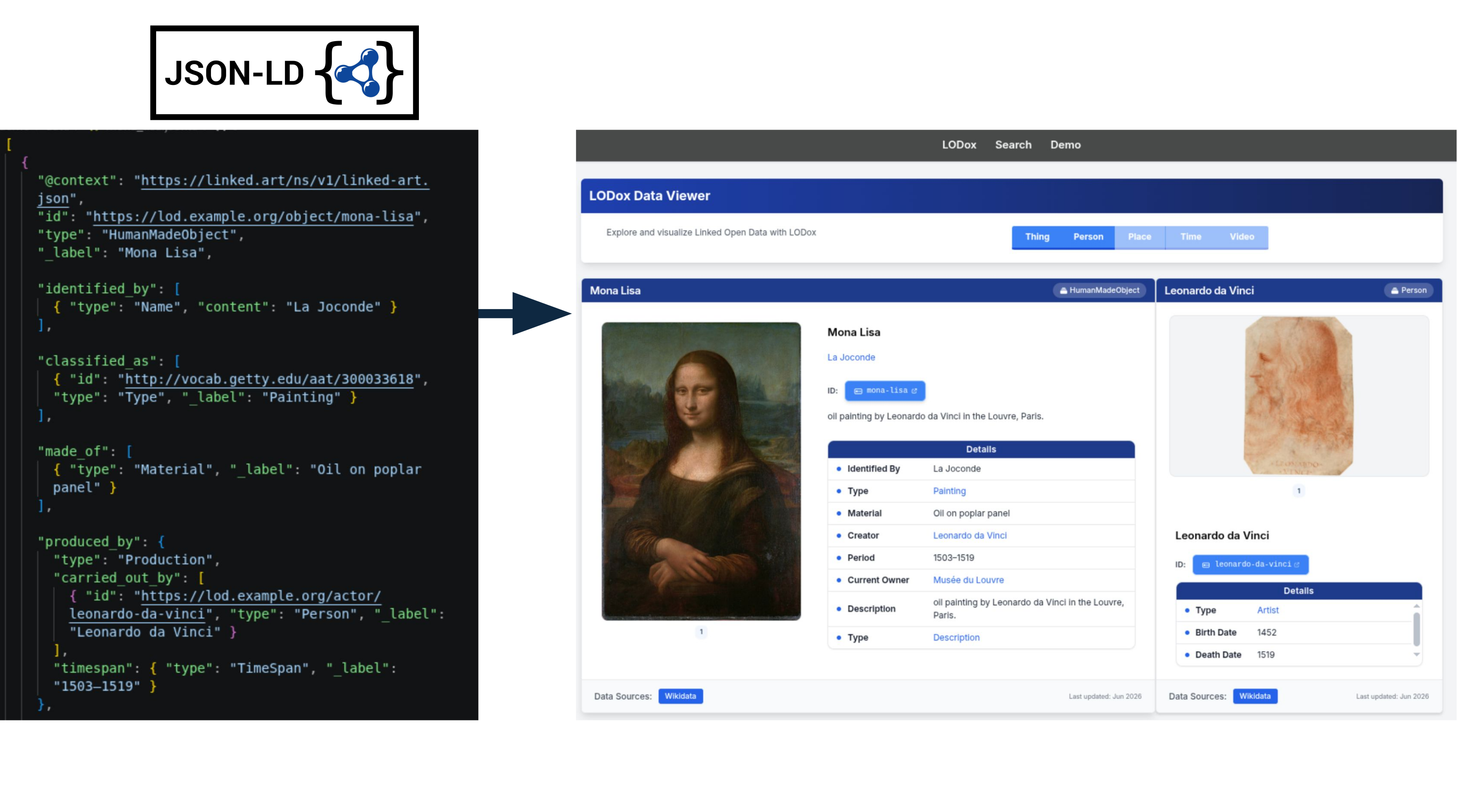}
    \caption{LODox Data Viewer, a Card-Based Data Viewer System.}
    \label{fig:card-viewer}
\end{figure}
LODox features a card-based visualization approach. It does not render data or graph data into tables or diagrams altogether. Key entities, or selected entities, such as cultural objects or persons, are displayed as cards with information inside. A page can feature multiple cards, and cards should have relationship with another card in the page. 

The card also has configurable layouts and sizes. A complex graph can then be visualized with several cards interacting with each other while each card presents name, descriptions, images and details. This is a scalable, entity (card)-centric alternative to traditional graph visualization. It is more suitable for cultural heritage domain where descriptive and narrative information is as important as relationships. This scalable approach is capable of presenting both.

On the card, the title (name) of the entity is on top, while its class (or classification) is on the upper-right corner. The bottom can used to present provenance information, for example, the data source.  Images may be placed either on the left side or in the upper half of the card, depending on the chosen customization. The name, ID, and description will be position accordingly automatically depending on the image size and availability.  Finally, a table is used to present additional details and its size is also configurable. For example, see Figure~\ref{fig:card-viewer}. 



\section{Solid-based Decentralized Ecosystem Vision}

The current approach, inheriting common practices from the field, is heavily centralized. It relies on museum experts to manually select terms from different vocabularies contributed by different sources, including users (implicit from their past queries), communities, and authoritative resources. This vocabulary becomes the only choice for users in their queries for valid results.

\begin{figure}
    \centering
    \includegraphics[width=1\linewidth]{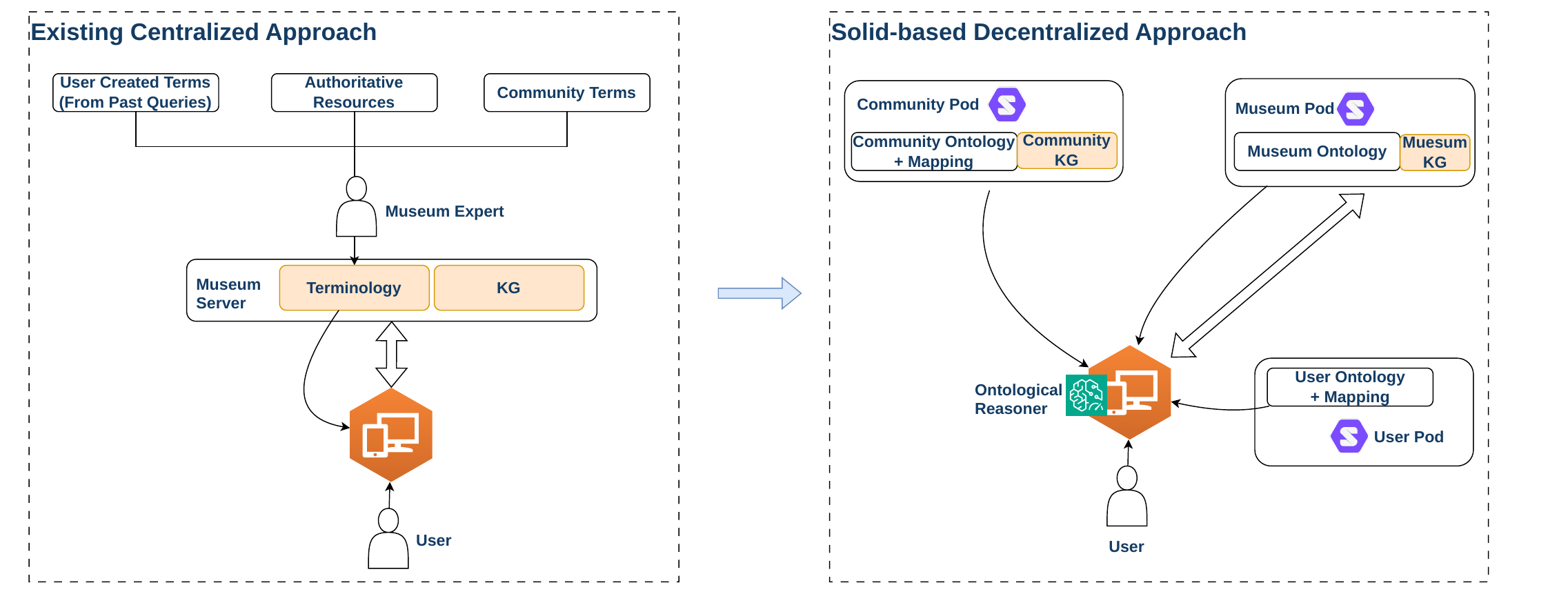}
    \caption{Visionary migration from current centralized approach to Solid-based decentralized approach}
    \label{fig:current-and-solid-approach}
\end{figure}

Solid allows a more decentralized approach, as illustrated in Fig. \ref{fig:current-and-solid-approach}, where users, museums and communities become more equal. Museums can still define its own vocabulary and have their KG. The community can also define their own vocabulary and the corresponding KG, separate from the museum's; the only special requirement is that they need to define mappings of their vocabularies to the museum's, thus the vocabulary is essentially an ontology. The user can choose any vocabulary to use, or define their own ontology, to query the museum's KG. All the differences and conversions between vocabularies are automatically handled by the ontological reasoner behind the scene. Solid\footnote{This requires the Solid server to have access-controlled SPARQL endpoint deployed together.} becomes the permission and cooperation layer in the ecosystem: the community or the museum stores its respective vocabulary and KG into its Solid Pod, and can authorize any other users to query (read) their data, or manage (write) or contribute (append) to their data; the identity is Web-scale, so they do not need to maintain separate user registries. This provides a neat solution and hopefully a healthier ecosystem, where everyone can silently speak different vocabularies while understanding each other, and agilely collaborate without unnecessary burden.

\section{Conclusion and Future Work}
In this paper, we presented LODox, a pragmatic approach to building accessible and secure exploration tools for cultural heritage Linked Open Data. 
LODox platform, designed as several modular, reusable components rather than tightly coupled system, is easier for future extensions. We also briefly discussed our vision for a better ecosystem based on Solid.

Our tech stack choices are also informed by the vision of integrating LLMs with Linked Data, which is why we store data as Linked Data while leveraging a property graph. We did not discuss that in details due to the lack of evaluation, further rooted in the small dataset.
Another limitation of this work is the absence of a formal user study, due to the rapid development at the moment with breaking changes frequently. 

Therefore, our future work will focus on three directions. First, we will integrate further components, such as additional LLMs agents, and work toward realising the pathway outlined in the vision. Second, we will also run experiments with larger datasets. we aim to reach a stable release of the  and release the source code as soon as possible. Third, that stable release will enable the formal user studies that rapid development has so far precluded.

\begin{acknowledgments}
The authors are grateful to Professor Donna Kurtz for her guidance and support in designing the LODox system.

\end{acknowledgments}

\section*{Declaration on Generative AI}

 During the preparation of this work, the author(s) used Claude and ChatGPT in order to: grammar and spelling check, content enhancement, improve writing style, and paraphrase and reword. After using these tools/services, the authors reviewed and edited the content as needed and take full responsibility for the publication’s content. 

\bibliography{bibfile}

@incollection{vrandecic_reshaping_2018,
	address = {Cham},
	title = {Reshaping the {Knowledge} {Graph} by {Connecting} {Researchers}, {Data} and {Practices} in {ResearchSpace}},
	volume = {11137},
	isbn = {978-3-030-00667-9 978-3-030-00668-6},
	url = {https://link.springer.com/10.1007/978-3-030-00668-6_20},
	language = {en},
	urldate = {2023-10-15},
	booktitle = {The {Semantic} {Web} – {ISWC} 2018},
	publisher = {Springer International Publishing},
	author = {Oldman, Dominic and Tanase, Diana},
	editor = {Vrandečić, Denny and Bontcheva, Kalina and Suárez-Figueroa, Mari Carmen and Presutti, Valentina and Celino, Irene and Sabou, Marta and Kaffee, Lucie-Aimée and Simperl, Elena},
	year = {2018},
	doi = {10.1007/978-3-030-00668-6_20},
	note = {Series Title: Lecture Notes in Computer Science},
	pages = {325--340},
}

@inproceedings{miranda_cidoc-crm_2025,
	address = {Portoroz, Slovenia},
	series = {{CEUR} {Workshop} {Proceedings}},
	title = {{CIDOC}-{CRM} and the {First} {Prototype} of a {Semantic} {Portal} for the {CHExRISH} project},
	volume = {4009},
	url = {https://ceur-ws.org/Vol-4009/#paper_14},
	language = {en},
	urldate = {2025-09-28},
	booktitle = {Proceedings of the {Second} {International} {Workshop} of {Semantic} {Digital} {Humanities} ({SemDH} 2025)},
	publisher = {CEUR},
	author = {Miranda, Luiz Do Valle and Kutt, Krzysztof and Nalepa, Grzegorz J.},
	editor = {Bruns, Oleksandra and Graciotti, Arianna and Sartini, Bruno and Tietz, Tabea},
	month = jun,
	year = {2025},
	note = {ISSN: 1613-0073},
}

@inproceedings{hyvonen_how_2025,
	address = {Portoroz, Slovenia},
	series = {{CEUR} {Workshop} {Proceedings}},
	title = {How to {Create} a {Portal} for {Digital} {Humanities} {Research} {Using} a {Linked} {Open} {Data} {Cloud} of {Cultural} {Heritage} {Knowledge} {Graphs}: {Case} {SampoSampo}},
	volume = {4009},
	shorttitle = {How to {Create} a {Portal} for {Digital} {Humanities} {Research} {Using} a {Linked} {Open} {Data} {Cloud} of {Cultural} {Heritage} {Knowledge} {Graphs}},
	url = {https://ceur-ws.org/Vol-4009/#paper_11},
	language = {en},
	urldate = {2025-09-30},
	booktitle = {Proceedings of the {Second} {International} {Workshop} of {Semantic} {Digital} {Humanities} ({SemDH} 2025)},
	publisher = {CEUR},
	author = {Hyvönen, Eero and Leskinen, Petri and Ahola, Annastiina and Rantala, Heikki and Tuominen, Jouni},
	editor = {Bruns, Oleksandra and Graciotti, Arianna and Sartini, Bruno and Tietz, Tabea},
	month = jun,
	year = {2025},
	note = {ISSN: 1613-0073},

}

@incollection{hutchison_linked-data-driven_2013,
	address = {Berlin, Heidelberg},
	title = {A {Linked}-{Data}-{Driven} and {Semantically}-{Enabled} {Journal} {Portal} for {Scientometrics}},
	volume = {7908},
	isbn = {978-3-642-38708-1 978-3-642-38709-8},
	url = {http://link.springer.com/10.1007/978-3-642-41338-4_8},
	urldate = {2025-10-02},
	booktitle = {Advanced {Information} {Systems} {Engineering}},
	publisher = {Springer Berlin Heidelberg},
	author = {Hu, Yingjie and Janowicz, Krzysztof and McKenzie, Grant and Sengupta, Kunal and Hitzler, Pascal},
	editor = {Hutchison, David and Kanade, Takeo and Kittler, Josef and Kleinberg, Jon M. and Mattern, Friedemann and Mitchell, John C. and Naor, Moni and Nierstrasz, Oscar and Pandu Rangan, C. and Steffen, Bernhard and Sudan, Madhu and Terzopoulos, Demetri and Tygar, Doug and Vardi, Moshe Y. and Weikum, Gerhard and Salinesi, Camille and Norrie, Moira C. and Pastor, \'Oscar},
	year = {2013},
	doi = {10.1007/978-3-642-41338-4_8},
	note = {Series Title: Lecture Notes in Computer Science},
	pages = {114--129},
}

@article{hyvonen_digital_2023,
	title = {Digital humanities on the {Semantic} {Web}: {Sampo} model and portal series},
	volume = {14},
	copyright = {https://creativecommons.org/licenses/by/4.0/},
	issn = {22104968, 15700844},
	shorttitle = {Digital humanities on the {Semantic} {Web}},
	url = {https://journals.sagepub.com/doi/full/10.3233/SW-223034},
	doi = {10.3233/SW-223034},
	number = {4},
	urldate = {2025-10-02},
	journal = {Semantic Web},
	author = {Hyvönen, Eero},
	editor = {Schlieder, Christoph},
	month = apr,
	year = {2023},
	pages = {729--744},
}

@article{ikkala_sampo-ui_2021,
	title = {Sampo-{UI}: {A} full stack {JavaScript} framework for developing semantic portal user interfaces},
	volume = {13},
	copyright = {https://creativecommons.org/licenses/by/4.0/},
	issn = {22104968, 15700844},
	shorttitle = {Sampo-{UI}},
	url = {https://journals.sagepub.com/doi/full/10.3233/SW-210428},
	doi = {10.3233/SW-210428},
	number = {1},
	urldate = {2025-10-02},
	journal = {Semantic Web},
	author = {Ikkala, Esko and Hyvönen, Eero and Rantala, Heikki and Koho, Mikko},
	editor = {Tudorache, Tania},
	month = nov,
	year = {2021},
	pages = {69--84},
}

@article{haase_metaphactory_2019,
	title = {metaphactory: {A} platform for knowledge graph management},
	volume = {10},
	issn = {22104968, 15700844},
	shorttitle = {metaphactory},
	url = {https://journals.sagepub.com/doi/full/10.3233/SW-190360},
	doi = {10.3233/SW-190360},
	number = {6},
	urldate = {2025-10-02},
	journal = {Semantic Web},
	author = {Haase, Peter and Herzig, Daniel M. and Kozlov, Artem and Nikolov, Andriy and Trame, Johannes},
	editor = {Kejriwal, Mayank and Lopez, Vanessa and Sequeda, Juan F. and Kejriwal, Mayank and Lopez, Vanessa and Sequeda, Juan F.},
	month = oct,
	year = {2019},
	pages = {1109--1125},
}

@misc{diallo_comprehensive_2024,
	title = {A {Comprehensive} {Evaluation} of {Neural} {SPARQL} {Query} {Generation} from {Natural} {Language} {Questions}},
	url = {http://arxiv.org/abs/2304.07772},
	doi = {10.48550/arXiv.2304.07772},
	urldate = {2025-10-02},
	publisher = {arXiv},
	author = {Diallo, Papa Abdou Karim Karou and Reyd, Samuel and Zouaq, Amal},
	month = jan,
	year = {2024},
	note = {arXiv:2304.07772 [cs]},
}

@misc{minaee_large_2025,
	title = {Large {Language} {Models}: {A} {Survey}},
	url = {https://arxiv.org/abs/2402.06196},
	author = {Minaee, Shervin and Mikolov, Tomas and Nikzad, Narjes and Chenaghlu, Meysam and Socher, Richard and Amatriain, Xavier and Gao, Jianfeng},
	year = {2025},
	note = {\_eprint: 2402.06196},
}

@inproceedings{hoseini_towards_2024,
	address = {Hersonissos, Greece},
	series = {{CEUR} {Workshop} {Proceedings}},
	title = {Towards {LLM}-augmented {Creation} of {Semantic} {Models} for {Dataspaces}},
	volume = {3705},
	url = {https://ceur-ws.org/Vol-3705/#paper03},
	language = {en},
	urldate = {2025-10-02},
	booktitle = {Proceedings of the {Second} {International} {Workshop} on {Semantics} in {Dataspaces} ({SDS} 2024)},
	publisher = {CEUR},
	author = {Hoseini, Sayed and Burgdorf, Andreas and Paulus, Alexander and Meisen, Tobias and Quix, Christoph and Pomp, André},
	editor = {Theissen-Lipp, Johannes and Colpaert, Pieter and Sowe, Sulayman K. and Curry, Edward and Decker, Stefan},
	month = may,
	year = {2024},
	note = {ISSN: 1613-0073},
	
}

@misc{fondi_integrating_2024,
	title = {Integrating {SPARQL} and {LLMs} for {Question} {Answering} over {Scholarly} {Data} {Sources}},
	url = {http://arxiv.org/abs/2409.18969},
	doi = {10.48550/arXiv.2409.18969},
	urldate = {2025-10-02},
	publisher = {arXiv},
	author = {Fondi, Fomubad Borista and Fidel, Azanzi Jiomekong and Camara, Gaoussou},
	month = nov,
	year = {2024},
	note = {arXiv:2409.18969 [cs]},
}

@misc{sambra_solid_2016,
	title = {Solid: a platform for decentralized social applications based on linked data},
	url = {http://emansour.com/research/meccano/solid_protocols.pdf},
	shorttitle = {Solid},
	publisher = {{MIT} {CSAIL} \& Qatar Computing Research Institute, Tech. Rep},
	author = {Sambra, Andrei Vlad and Mansour, Essam and Hawke, Sandro and Zereba, Maged and Greco, Nicola and Ghanem, Abdurrahman and Zagidulin, Dmitri and Aboulnaga, Ashraf and Berners-Lee, Tim},
	urldate = {2026-06-06},
	date = {2016},
	year = {2016}
}


\end{document}